\documentstyle[epsfig,12pt,a4p]{article}

\newcommand{\pipi}{\mbox{$\pi^+\pi^-$ }}

\begin{document}
\begin{titlepage}
\def\footnoterule{\hrule width 1.0\columnwidth}
%\hfill  \hfill
%6\thinspace February\thinspace 1990
% \begin{center} {\large EUROPEAN ORGANIZATION FOR NUCLEAR RESEARCH}
%  \end{center}
\begin{tabbing}
put this on the right hand corner using tabbing so it looks
 and neat and in \= \kill
\> {10 February 1999}
\end{tabbing}
\bigskip
\bigskip
\begin{center}{\Large  {\bf A partial wave analysis of the centrally produced
$\pi^+ \pi^-$ system
in $pp$ interactions at 450 GeV/c}
}\end{center}
\bigskip
\bigskip
\begin{center}{        The WA102 Collaboration
}\end{center}\bigskip
\begin{center}{
D.\thinspace Barberis$^{  4}$,
W.\thinspace Beusch$^{   4}$,
F.G.\thinspace Binon$^{   6}$,
A.M.\thinspace Blick$^{   5}$,
F.E.\thinspace Close$^{  3,4}$,
K.M.\thinspace Danielsen$^{ 10}$,
A.V.\thinspace Dolgopolov$^{  5}$,
S.V.\thinspace Donskov$^{  5}$,
B.C.\thinspace Earl$^{  3}$,
D.\thinspace Evans$^{  3}$,
B.R.\thinspace French$^{  4}$,
T.\thinspace Hino$^{ 11}$,
S.\thinspace Inaba$^{   8}$,
A.V.\thinspace Inyakin$^{  5}$,
T.\thinspace Ishida$^{   8}$,
A.\thinspace Jacholkowski$^{   4}$,
T.\thinspace Jacobsen$^{  10}$,
G.T\thinspace Jones$^{  3}$,
G.V.\thinspace Khaustov$^{  5}$,
T.\thinspace Kinashi$^{  12}$,
J.B.\thinspace Kinson$^{   3}$,
A.\thinspace Kirk$^{   3}$,
W.\thinspace Klempt$^{  4}$,
V.\thinspace Kolosov$^{  5}$,
A.A.\thinspace Kondashov$^{  5}$,
A.A.\thinspace Lednev$^{  5}$,
V.\thinspace Lenti$^{  4}$,
S.\thinspace Maljukov$^{   7}$,
P.\thinspace Martinengo$^{   4}$,
I.\thinspace Minashvili$^{   7}$,
% K.\thinspace Myklebost$^{   3}$,
T.\thinspace Nakagawa$^{  11}$,
K.L.\thinspace Norman$^{   3}$,
% J.M.\thinspace Olsen$^{   3}$,
J.P.\thinspace Peigneux$^{  1}$,
S.A.\thinspace Polovnikov$^{  5}$,
V.A.\thinspace Polyakov$^{  5}$,
%%Yu.D.\thinspace Prokoshkin$^{\dag  5}$,
V.\thinspace Romanovsky$^{   7}$,
H.\thinspace Rotscheidt$^{   4}$,
V.\thinspace Rumyantsev$^{   7}$,
N.\thinspace Russakovich$^{   7}$,
V.D.\thinspace Samoylenko$^{  5}$,
A.\thinspace Semenov$^{   7}$,
M.\thinspace Sen\'{e}$^{   4}$,
R.\thinspace Sen\'{e}$^{   4}$,
P.M.\thinspace Shagin$^{  5}$,
H.\thinspace Shimizu$^{ 12}$,
A.V.\thinspace Singovsky$^{ 1,5}$,
A.\thinspace Sobol$^{   5}$,
A.\thinspace Solovjev$^{   7}$,
M.\thinspace Stassinaki$^{   2}$,
J.P.\thinspace Stroot$^{  6}$,
V.P.\thinspace Sugonyaev$^{  5}$,
K.\thinspace Takamatsu$^{ 9}$,
G.\thinspace Tchlatchidze$^{   7}$,
T.\thinspace Tsuru$^{   8}$,
%%G.\thinspace Vassiliadis$^{\dag   2}$,
M.\thinspace Venables$^{  3}$,
O.\thinspace Villalobos Baillie$^{   3}$,
M.F.\thinspace Votruba$^{   3}$,
Y.\thinspace Yasu$^{   8}$.
% \end authorlist
}\end{center}

\begin{center}{\bf {{\bf Abstract}}}\end{center}

{
A partial wave analysis of the centrally produced \pipi
channel has been performed in $pp$ collisions using
an incident beam momentum of 450~GeV/c.
An unambiguous physical solution has been found.
Evidence is found for the $f_0(980)$, $f_0(1300)$, $f_0(1500)$
and $f_J(1710)$ with J~=~0 in the
the S-wave.
The $\rho(770)$ is observed dominantly in the $P_0^-$-wave
and the $f_2(1270)$ is observed dominantly in the $D_0^-$-wave.
In addition, there is evidence for a
broad enhancement in the D-wave below 1 GeV.
}
\bigskip
\bigskip
\bigskip
\bigskip\begin{center}{{Submitted to Physics Letters}}
\end{center}
%\newpage
\bigskip
\bigskip
\begin{tabbing}
aba \=   \kill
% $^\dag$ \> \small
% Deceased. \\
$^1$ \> \small
LAPP-IN2P3, Annecy, France. \\
$^2$ \> \small
Athens University, Physics Department, Athens, Greece. \\
% $^3$ \> \small
% Bergen University, Bergen, Norway. \\
$^3$ \> \small
School of Physics and Astronomy, University of Birmingham, Birmingham, U.K. \\
$^4$ \> \small
CERN - European Organization for Nuclear Research, Geneva, Switzerland. \\
$^5$ \> \small
IHEP, Protvino, Russia. \\
$^6$ \> \small
IISN, Belgium. \\
$^7$ \> \small
JINR, Dubna, Russia. \\
$^8$ \> \small
High Energy Accelerator Research Organization (KEK), Tsukuba, Ibaraki 305,
Japan. \\
$^{9}$ \> \small
Faculty of Engineering, Miyazaki University, Miyazaki, Japan. \\
$^{10}$ \> \small
Oslo University, Oslo, Norway. \\
$^{11}$ \> \small
Faculty of Science, Tohoku University, Aoba-ku, Sendai 980, Japan. \\
$^{12}$ \> \small
Faculty of Science, Yamagata University, Yamagata 990, Japan. \\
\end{tabbing}
\end{titlepage}
\setcounter{page}{2}
\bigskip
\par
Lattice gauge calculations
indicate that the lowest lying glueball should be a scalar and be in the
mass range 1500-1700~MeV~\cite{re:lgt}.
However,
after more than 20 years of searching for glueballs there is emerging
evidence that a pure scalar glueball may never be observed due to the fact
that its finite width means that it mixes with nearby isoscalar
$q \overline q$ states with $J^{PC}$~=~$0^{++}$~\cite{re:cfl}.
This scenario
gives rise to several states whose properties are complicated and
lie somewhere between conventional $q \overline q$ states and glueballs.
Hence a systematic study of all isoscalar
$J^{PC}$~=~$0^{++}$ states is required.
The $\pi \pi$ system has long been a profitable
place to study $I^GJ^{PC}$~=~$0^+0^{++}$ states and to date
very high statistics studies have been performed using $\pi$ induced
interactions~\cite{piin} and $p \overline p$ annihilations~\cite{ppbar}.
This paper presents a study of the
\pipi
final state formed in central $pp$ interactions which
are predicted to be a source of gluonic final states
via double Pomeron exchange~\cite{closerev}.
\par
The reaction
\begin{equation}
pp \rightarrow p_{f} (\pi^+ \pi^-) p_{s}
\label{eq:b}
\end{equation}
has been studied at 450~GeV/c.
The subscripts $f$ and $s$ indicate the
fastest and slowest particles in the laboratory respectively.
The WA102 experiment
has been performed using the CERN Omega Spectrometer,
the layout of which is
described in ref.~\cite{WADPT}.
Reaction (1)
has been isolated
from the sample of events having four
outgoing
charged tracks,
by first imposing the following cuts on the components of the
missing momentum:
$|$missing~$P_{x}| <  14.0$ GeV/c,
$|$missing~$P_{y}| <  0.16$ GeV/c and
$|$missing~$P_{z}| <  0.08$ GeV/c,
where the x axis is along the beam
direction.
A correlation between
pulse-height and momentum
obtained from a system of
scintillation counters was used to ensure that the slow
particle was a proton.
\par
The method of Ehrlich et al.~\cite{EHRLICH},
has been used to compute the mass
squared of the two centrally produced particles assuming them to have
equal mass.
The resulting distribution is shown in
fig.~\ref{fi:1}a) where a clear peak can be seen at
the pion mass squared.
A cut on the Ehrlich mass squared of
$-0.3 \leq M^2_X \leq 0.2$~$GeV^2$
has been used to select a sample of 5.15 million \pipi events.
\par
Fig.~\ref{fi:1}b) shows
the $p_f \pi^+$ effective mass spectrum where
a clear $\Delta^{++}(1232)$
can be observed.
A smaller $\Delta^0(1232)$ signal can be seen in the
$p_f \pi^-$ effective mass spectrum shown in fig.~\ref{fi:1}c).
The $\Delta(1232)$ signal has been removed
by requiring $M(p_f\pi)$~$>$~1.5~GeV.
Due to the trigger requirements there
is little evidence for $\Delta$ production in the
$p_s\pi$ effective mass spectra (not shown). However, for symmetry purposes
a cut of
$M(p_s\pi)$~$>$~1.5~GeV has also been applied.
\par
As can be seen from
fig.~\ref{fi:1}b)
there is also some evidence for higher mass $\Delta^{++}$ or $N^*$ production.
In order to investigate the effect of this on the
following results an analysis has also been performed
requiring that $M(p\pi)$~$>$~2.0 GeV. It is found that the
higher mass proton excitations do not significantly influence the results.
\par
The Feynman $x_F$ distributions for the slow particle,
the \pipi system and the fast particle
are shown in fig.~\ref{fi:1}d).
As can be seen the \pipi system lies within $|x_F| \leq 0.25$.
\par
The resulting centrally produced \pipi effective mass distribution is shown
in fig.~\ref{fi:1}e) and consists of 2.87 million events.
As can be seen there is
evidence for a small $\rho^0(770)$ signal, some $f_2(1270)$ and a
sharp drop at 1~GeV which has
been interpreted as being due to the interference of the
$f_0(980)$ with the S-wave background~\cite{oldpipi}.
\par
A Partial Wave Analysis (PWA) of the centrally produced \pipi system has been
performed assuming the \pipi system is produced by the
collision of two particles (referred to as exchanged particles) emitted
by the scattered protons.
The z axis
is defined by the momentum vector of the
exchanged particle with the greatest four-momentum transferred
in the \pipi centre of mass.
The y axis is defined
by the cross product of the two exchanged particles in the $pp$ centre of mass.
The two variables needed to specify the decay process were taken as the polar
and azimuthal angles ($\theta$, $\phi$) of the $\pi^-$ in the \pipi
centre of mass relative to the coordinate system described above.
\par
The acceptance corrected moments, defined by
\begin{equation}
I(\Omega) =  \sum _L t_{L0} Y^0 _L(\Omega) + 2 \sum _{L,M >0}
t_{LM}Re\{Y^M_L(\Omega)\}
\end{equation}
have been rescaled to the total number of observed events and
are shown
in fig.~\ref{fi:2}.
As can be seen the moments
with $M$~$>$~2 and $L$~$>$~4 are small (i.e. $t_{43}$, $t_{44}$,
$t_{50}$ and $t_{60}$)
and hence only
S, P, and D waves with $m$~$\leq$~1 have been included in the PWA.
\par
An interesting feature of the moments is the presence of a structure
in the $L$~=~4 moments for \pipi masses below 1~GeV which
indicates the presence of D-wave.
This type of structure has not been observed in $\pi$ induced
reactions~\cite{piin}.
In order to see if this effect is due to acceptance problems or
problems due to non-central events, we have reanalysed the data
using a series of different cuts.
Firstly, we required that $M(p\pi)$~$>$2.0~GeV; however, after acceptance
correction the moments were compatible with the set for
$M(p\pi)$~$>$1.5~GeV showing that diffractive resonances in the range
1.5~$<$~$M(p\pi)$~$<$~2.0~GeV have a negligible effect on the moments.
We have also required that the rapidity gap between any proton and
$\pi$ in the event is greater than 2 units. Again the resulting
acceptance corrected moments do not change.
In order to investigate any systematic effects we have also analysed the
central $\pi^0\pi^0$ data and a similar structure is also found
in the $L$~=~4 moments~\cite{WA102pi0pi0}.
\par
It is interesting to compare the above result with
previously published data on other centrally produced $\pi \pi$ systems.
In preliminary results from the E690 experiment
at Fermilab activity at a similar level is observed in the $t_{40}$ moment
below 1~GeV~\cite{E690pipi}.
There is also evidence for this structure in the $\pi^0\pi^0$
data from the NA12/2 experiment~\cite{NA122pi0pi0}.
The AFS experiment at the CERN
ISR also observed that the $t_{40}$ moments deviated from zero
in this mass region; however, in their analysis they
claimed this deviation was due to problems of the Monte Carlo simulating
low energy tracks~\cite{AFS}.
\par
This structure does indeed seem to be a real effect
which is present in centrally produced $\pi \pi$ systems.
It has recently been suggested~\cite{fec,jmf}
that central production may be due to the
fusion of two vector particles and this may explain why
higher angular momentum systems can be produced at lower masses.
\par
The amplitudes used for the PWA are defined in the reflectivity
basis~\cite{reflectivity}.
In this basis the angular distribution is given by a sum of two non-interfering
terms corresponding to negative and positive values of reflectivity.
The waves used were of the form $J^\varepsilon _m$ with $J$~$=$~S, P and D,
$m$~$=$~$0,1$ and reflectivity $\varepsilon$~=~$\pm 1$.
The expressions relating the moments
($t_{LM}$) and the waves ($J^\varepsilon _m$) are given in table~\ref{ta:a}.
Since the overall phase for each reflectivity is indeterminate,
one wave in each reflectivity can be set to be real ($S_0^-$ and $P_1^+$
for example) and hence two phases can be set to zero ($\phi_{S_0^-}$ and
$\phi_{P_1^+}$ have been chosen).
This results in 12 parameters to be determined from the fit to the
angular distributions.
% This results in 12 parameters to be determined from the 12 non-zero moments.
\par
The PWA has been performed independently in 20~MeV intervals of the \pipi
mass spectrum. In each mass an event-by-event maximum likelihood
method has been used. The function
\begin{equation}
F=-\sum_{i=1}^Nln\{I(\Omega)\} + \sum_{L,M}t_{LM}\epsilon_{LM}
\end{equation}
has been minimised, where N is the number of events in a given mass bin,
$\epsilon_{LM}$ are the efficiency corrections
calculated in the centre of the bin
and $t_{LM}$ are the moments of the angular distribution.
% The results of the fit are shown superimposed on the moments
% used in the fit in fig~\ref{fi:2}. As can be seen the results of the fit
The moments calculated from the partial amplitudes
are shown superimposed on the experimental moments
in fig~\ref{fi:2}. As can be seen the results of the fit
well reproduce the experimental moments.
\par
The system of equations which express the moments via the partial wave
amplitudes is non-linear that leads to inherent
ambiguities. For a system with S, P and D waves there are eight solutions
for each mass bin.
In each mass bin
one of these solutions
is found from the fit to the experimental angular distributions
while the other seven can then be calculated by the method described in
ref.~\cite{reflectivity}.
In order to link the solutions in adjacent mass bins, the real and
imaginary parts of the Barrelet function roots are required to be
step-wise
continuous and have finite derivatives as a function of mass~\cite{link}.
By definition, all the solutions give identical moments and identical
values of the likelihood.
The only way to differentiate between the solutions, if different, is to
apply some external physical test, such as requiring
that at threshold that the S-wave is the dominant wave.
\par
The four complex roots, Z$_i$, after the linking procedure are shown in
fig.~\ref{fi:3}.
As can be seen
the real parts are
well separated
and hence it is possible to identify unambiguously
all the PWA solutions in the whole mass range.
In addition,
the zeros do not cross the real axis
and hence there is no problem with bifurcation of the solutions.
Near threshold the P-wave is the dominant contribution for five
solutions, another one is dominated by D-wave and
another has the same amount of S-wave and P-wave. These seven solutions
have been ruled out because the \pipi cross section near threshold
has been assumed
to be dominated by S-wave. The remaining solution is shown in fig.~\ref{fi:4}.
\par
The S-wave spectrum shows a clear threshold enhancement followed by a sharp
drop at 1~GeV. There is clear evidence for the $\rho(770)$ in the $P_0^-$ wave
and for the $f_2(1270)$ in the
$D_0^-$ wave.
\par
In order to obtain a satisfactory fit
to the $S_0^-$ wave from threshold to 2~GeV it has been found to be
necessary to use
three interfering Breit-Wigners to describe the $f_0(980)$, $f_0(1300)$
and $f_0(1500)$ and a background
of the form
$a(m-m_{th})^{b}exp(-cm-dm^{2})$, where
$m$ is the
\pipi
mass,
$m_{th}$ is the
\pipi
threshold mass and
a, b, c, d are fit parameters.
The Breit-Wigners have been convoluted with a Gaussian to
account for the experimental mass resolution
($\sigma$~=~4 MeV at threshold rising to 22~MeV at 2~GeV).
The fit is shown in fig.~\ref{fi:3}c) for the entire mass range
and in fig.~\ref{fi:3}d) for masses above 1 GeV.
The resulting parameters are
\begin{tabbing}
asda\=adfsfsf99ba \=Mas \= == \=1224 \=pm \=1200 \=MeVswfw, \=gaa \=  == \=1224
\=pm \=1200  \=MeV   \kill
\>$f_0(980)$ \>M \>=\>982\>$\pm$\>3\>MeV,\>$\Gamma$\>=\>80\>$\pm$\>10\>MeV \\
\>$f_0(1300)$ \>M \>=\>1308\>$\pm$\>10\>MeV,\>$\Gamma$\>=\>222\>$\pm$\>20\>MeV
\\
\>$f_0(1500)$ \>M \>=\>1502\>$\pm$\>10\>MeV,\>$\Gamma$\>=\>131\>$\pm$\>15\>MeV
\end{tabbing}
which are consistent with the PDG~\cite{PDG98} values for these
resonances.
As can be seen, the fit describes the data well for masses below 1~GeV.
It was not possible to describe the data above 1~GeV without the addition
of both the $f_0(1300)$ and $f_0(1500)$ resonances.
However, even with this fit using
three Breit-Wigners it can be seen that the fit does not
describe well the 1.7 GeV region.
This could be due to a \pipi decay mode of the $f_J(1710)$ with J~=~0.
Including a fourth Breit-Wigner in this mass region decreases the
$\chi^2$ from 256 to 203 and yields
\begin{tabbing}
asda\=adfsfsf99ba \=Mas \= == \=1224 \=pm \=1200 \=MeVswfw, \=gaa \=  == \=1224
\=pm \=1200  \=MeV   \kill
\>$f_J(1710)$ \>M \>=\>1750\>$\pm$\>20\>MeV,\>$\Gamma$\>=\>160\>$\pm$\>30\>MeV.
\end{tabbing}
parameters which are consistent with the PDG~\cite{PDG98} values for the
$f_J(1710)$.
The fit is
shown in fig.~\ref{fi:3}e) for masses above 1 GeV.
\par
In conclusion, a partial wave analysis of a
high statistics sample of centrally produced \pipi events has been performed.
An unambiguous physical solution has been found.
The S-wave is found to dominate the mass spectrum and is composed of
a broad enhancement at threshold, a sharp drop
at 1 GeV due to the interference between the $f_0(980)$
and the S-wave background, the $f_0(1300)$, the $f_0(1500)$ and
the $f_J(1710)$ with J~=~0.
The $\rho(770)$ is observed dominantly in the $P_0^-$-wave.
The D-wave shows evidence for the $f_2(1270)$ and a broad enhancement
below 1 GeV. It is interesting to note that the $f_2(1270)$ is
produced dominantly with m~=~0.
There is no evidence for any
significant structure in the D-wave above the $f_2(1270)$
i.e. no evidence for a J~=~2 component of the $f_J(1710)$.
\begin{center}
{\bf Acknowledgements}
\end{center}
\par
This work is supported, in part, by grants from
the British Particle Physics and Astronomy Research Council,
the British Royal Society,
the Ministry of Education, Science, Sports and Culture of Japan
(grants no. 04044159 and 07044098), the Programme International
de Cooperation Scientifique (grant no. 576)
and
the Russian Foundation for Basic Research
(grants 96-15-96633 and 98-02-22032).

\newpage
\newpage
\begin{table}[h]
\caption{The moments of the angular distribution expressed in terms of the
partial waves.}
\label{ta:a}
\vspace{1in}
\begin{center}
\begin{tabular}{|ccl|} \hline
  & & \\
 $\sqrt{4 \pi}t_{00}$ & = &
$|S_0^-|^2+|P_0^-|^2+|P^-_1|^2+|P^+_1|^2+|D_0^-|^2+|D^-_1|^2+|D^+_1|^2$\\
  & & \\
 $\sqrt{4 \pi}t_{10}$ & = &
$2|S_0^-||P_0^-|cos(\phi_{S_0^-}-\phi_{P_0^-})+
\frac{4}{\sqrt{5}}|P_0^-||D_0^-|cos(\phi_{P_0^-}-\phi_{D_0^-})$\\
&&$
+\frac{2\sqrt{3}}{\sqrt{5}}\{|P_1^-||D_1^-|cos(\phi_{P_1^-}-
\phi_{D_1^-})+|P_1^+||D_1^+|cos(\phi_{P_1^+}-\phi_{D_1^+})\}$\\
  & & \\
 $\sqrt{4 \pi}t_{11}$ & = &
$\sqrt{2}|S_0^-||P_1^-|cos(\phi_{S_0^-}-\phi_{P_1^-})-
\frac{\sqrt{2}}{\sqrt{5}}|P_1^-||D_0^-|cos(\phi_{P_1^-}-\phi_{D_0^-})$\\
&&$ +\frac{\sqrt{6}}{\sqrt{5}}|P_0^-||D_1^-|cos(\phi_{P_0^-}-\phi_{D_1^-})$\\
  & & \\
 $\sqrt{4 \pi}t_{20}$ & = &
$\frac{2}{\sqrt{5}}|P_0^-|^2-\frac{1}{\sqrt{5}}(|P_1^-|^2+|P_1^+|^2)
+\frac{\sqrt{5}}{7}(2|D_0^-|^2 + |D_1^-|^2 + |D_1^+|^2)$\\
&&$ +2|S_0^-||D_0^-|cos(\phi_{S_0^-}-\phi_{D_0^-})$\\
  & & \\
 $\sqrt{4 \pi}t_{21}$ & = &
$\frac{\sqrt{6}}{\sqrt{5}}|P_1^-||P_0^-|cos(\phi_{P_1^-}-
\phi_{P_0^-})+\frac{\sqrt{10}}{7}|D_1^-||D_0^-
|cos(\phi_{D_1^-}-\phi_{D_0^-})$\\
&&$ +\sqrt{2}|S_0^-||D_1^-|cos(\phi_{S_0^-}-\phi_{D_1^-})$\\
  & & \\
 $\sqrt{4 \pi}t_{22}$ & = &
$\frac{\sqrt{3}}{\sqrt{10}}(|P_1^-|^2-|P_1^+|^2)+
\frac{\sqrt{15}}{7\sqrt{2}}(|D_1^-|^2-|D_1^+|^2)$\\
  & & \\
 $\sqrt{4 \pi}t_{30}$ & = &$
-\frac{6}{\sqrt{35}}\{|P_1^-||D_1^-|cos(\phi_{P_1^-}-
\phi_{D_1^-})+|P_1^+||D_1^+|cos(\phi_{P_1^+}-\phi_{D_1^+})\}$\\
 &&
$+\frac{6\sqrt{3}}{\sqrt{35}}|P_0^-||D_0^-|cos(\phi_{P_0^-}-\phi_{D_0^-})$\\
  & & \\
 $\sqrt{4 \pi}t_{31}$ & = &
$\frac{6}{\sqrt{35}}|P_1^-||D_0^-|cos(\phi_{P_1^-}-
\phi_{D_0^-})+\frac{4\sqrt{3}}{\sqrt{35}}|P_0^-
|D_1^-|cos(\phi_{P_0^-}-\phi_{D_1^-})$\\
  & & \\
 $\sqrt{4 \pi}t_{32}$ & = &
$\frac{\sqrt{6}}{\sqrt{7}}\{|P_1^-||D_1^-|cos(\phi_{P_1^-}-
\phi_{D_1^-})-|P_1^+||D_1^+|cos(\phi_{P_1^+}-\phi_{D_1^+})\}$\\
  & & \\
 $\sqrt{4 \pi}t_{40}$ & = &
$\frac{6}{7}|D_0^-|^2-\frac{4}{7}(|D_1^-|^2+|D_1^+|^2)$\\
  & & \\
 $\sqrt{4 \pi}t_{41}$ & = &
$\frac{2\sqrt{15}}{7}|D_0^-||D_1^-|cos(\phi_{D_0^-}-\phi_{D_1^-})$\\
  & & \\
 $\sqrt{4 \pi}t_{42}$ & = & $\frac{\sqrt{10}}{7}(|D_1^-|^2-|D_1^+|^2)$\\
  & & \\ \hline
\end{tabular}
\end{center}
\end{table}
\clearpage
{ \large \bf Figures \rm}
\begin{figure}[h]
\caption{a) The Ehrlich mass squared distribution,
b) the $M(p_f\pi^+)$ and c) the $M(p_f\pi^-)$ mass spectra.
d) The $x_F$ distribution for the slow particle, the \pipi system and
the fast particle.
e) The centrally produced \pipi effective mass spectrum.
}
\label{fi:1}
\end{figure}
\begin{figure}[h]
\caption{ The $\protect\sqrt{4\pi}t_{LM}$ moments from the data.
Superimposed as a solid histogram are the resulting moments calculated
from the PWA of the
\pipi final state.
}
\label{fi:2}
\end{figure}
\begin{figure}[h]
\caption{
a) The Real and b) Imaginary parts of the roots (see text)
as a function of mass obtained from the PWA.
c) and d) The \pipi $S_0^-$ wave with fit described in the text using three
Breit-Wigners.
e) The \pipi $S_0^-$ wave with fit described in the text using four
Breit-Wigners.
}
\label{fi:3}
\end{figure}
\begin{figure}[h]
\caption{
The physical solution from the PWA of the \pipi final state.
}
\label{fi:4}
\end{figure}
\newpage
\begin{center}
\epsfig{figure=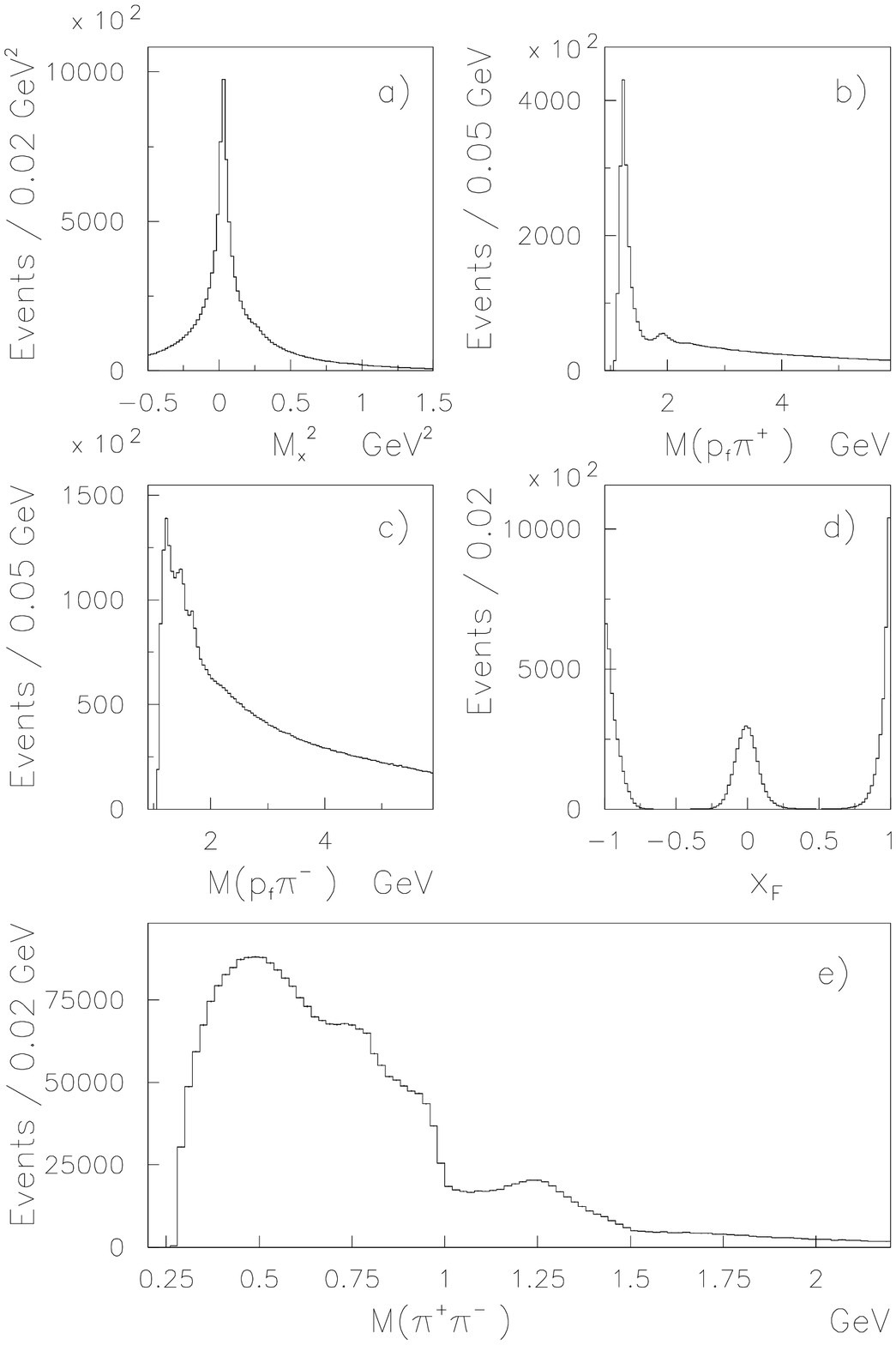,height=22cm,width=17cm}
\end{center}
\begin{center} {Figure 1} \end{center}
\newpage
\begin{center}
\epsfig{figure=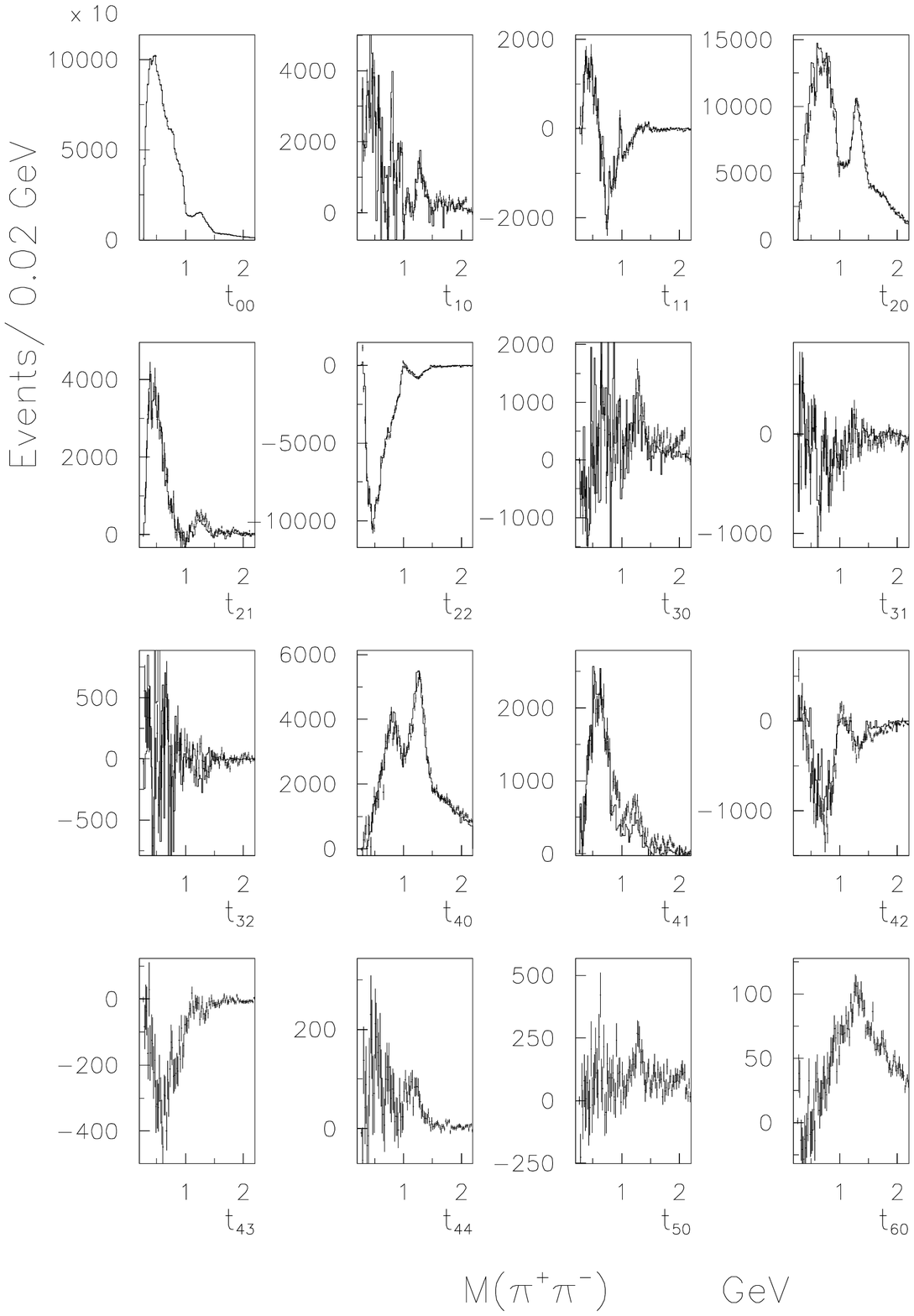,height=22cm,width=17cm,
bbllx=0pt,bblly=0pt,bburx=550pt,bbury=700pt}
\end{center}
\begin{center} {Figure 2} \end{center}
\newpage
\begin{center}
\epsfig{figure=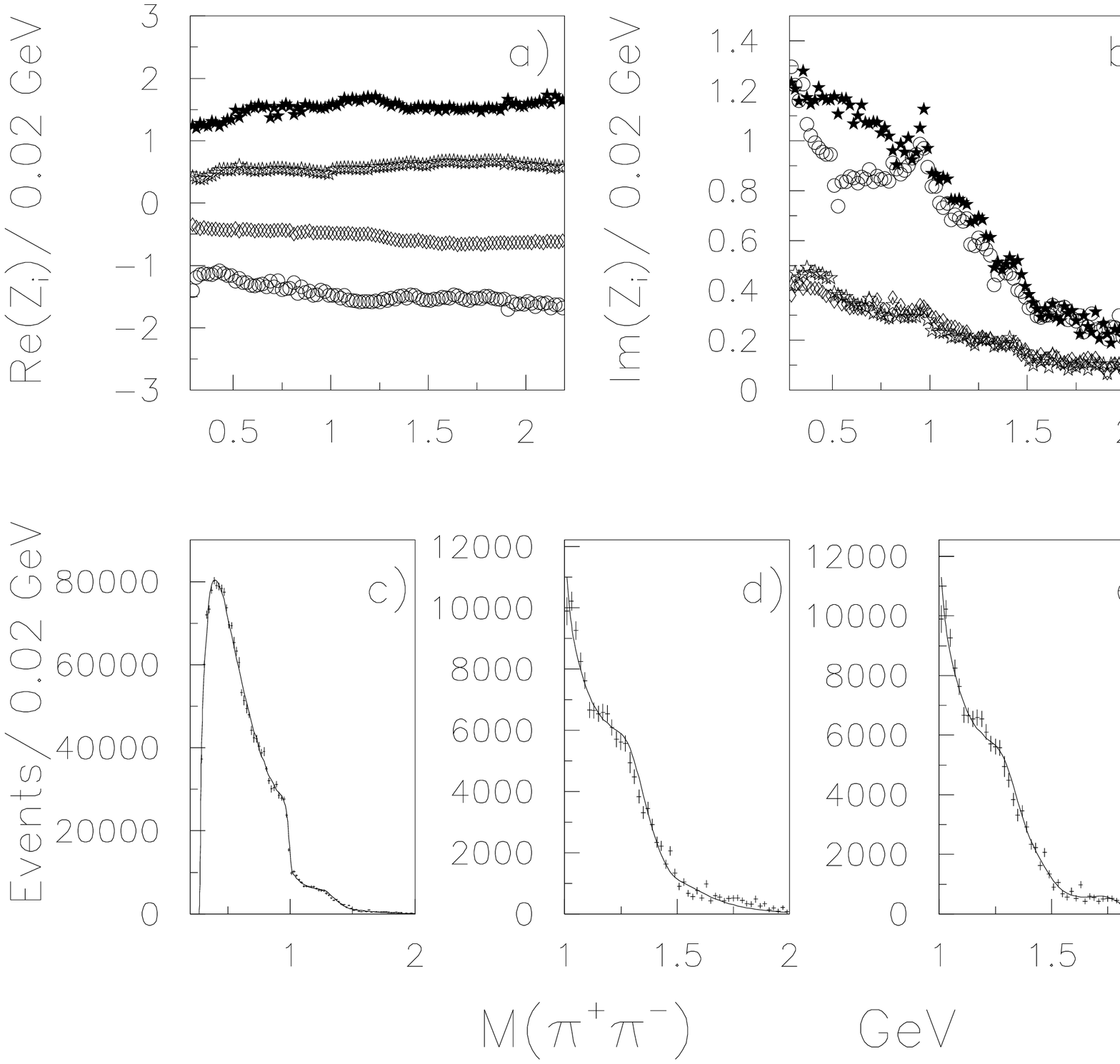,height=22cm,width=17cm,
bbllx=0pt,bblly=0pt,bburx=600pt,bbury=750pt}
\end{center}
\begin{center} {Figure 3} \end{center}
\newpage
\begin{center}
\epsfig{figure=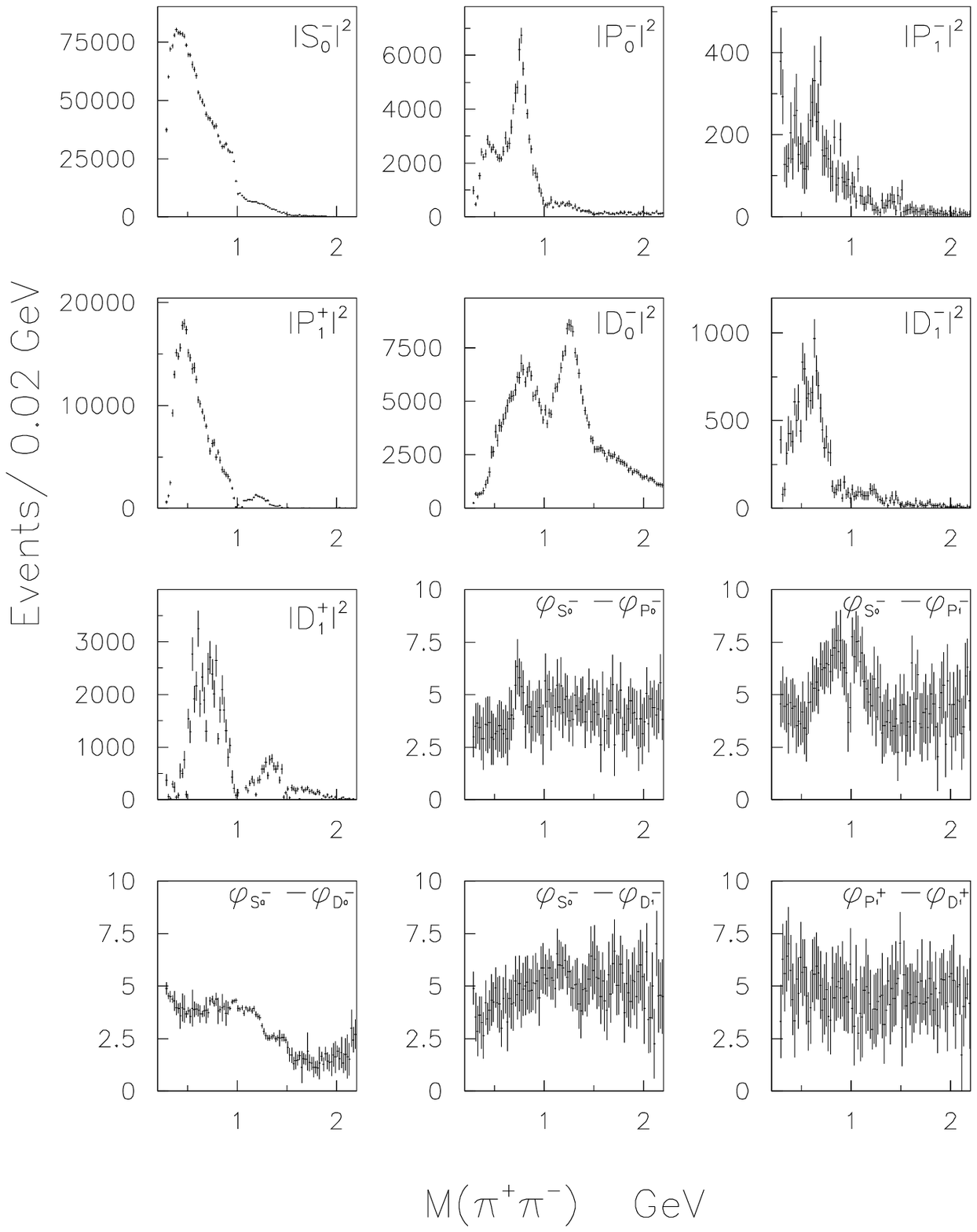,height=22cm,width=17cm,
bbllx=0pt,bblly=0pt,bburx=550pt,bbury=700pt}
\end{center}
\begin{center} {Figure 4} \end{center}
\end{document}